\documentclass[12pt]{article}
\usepackage{epsfig}

\parskip=\medskipamount 
plus.3em minus.5em \abovedisplayshortskip=.5em plus.2em minus.4em
\renewcommand{\thesection}{\arabic{section}}

\catcode`@=11

\@addtoreset{equation}{section} \@addtoreset{equation}{subsection}
\def\theequation{\ifnum\value{section}=0 \arabic{equation}\ignorespaces
\else \ifnum\value{section}=-1 A.\arabic{equation}\ignorespaces
\else \ifnum\value{subsection}=0
\thesection.\arabic{equation}\ignorespaces \else
\thesection.\arabic{subsection}.\arabic{equation}\ignorespaces
                             \fi
                        \fi
                   \fi}

{\catcode`\'=\active \def'{{}^\bgroup\prim@s}}

\catcode`@=12

\newcommand{\bq}{\begin{equation}}
\newcommand{\be}{\begin{equation}}
\newcommand{\fq}{\end{equation}}
\newcommand{\ee}{\end{equation}}
\newcommand{\bqr}{\begin{eqnarray}}
\newcommand{\beqs}{\begin{eqnarray}}
\newcommand{\fqr}{\end{eqnarray}}
\newcommand{\eeqs}{\end{eqnarray}}

\newcommand{\rf}[1]{(\ref{#1})}

\def\del{\delta}

\def\bop#1{\setbox0=\hbox{$#1M$}\mkern1.5mu
    \vbox{\hrule height0pt depth.04\ht0
    \hbox{\vrule width.04\ht0 height.9\ht0 \kern.9\ht0
    \vrule width.04\ht0}\hrule height.04\ht0}\mkern1.5mu}
\def\Box{{\mathpalette\bop{}}}                        

\begin{document}
\thispagestyle{empty}

\begin{flushright}
\begin{tabular}{l}
\end{tabular}
\end{flushright}

\vskip .6in
\begin{center}

{\Large\bf  Symmetry Algebra of IIB Superstring Scattering}

\vskip .6in

{\bf Gordon Chalmers}
\\[5mm]

{e-mail: gordon@quartz.shango.com}

\vskip .5in minus .2in

{\bf Abstract}

\end{center}

The graviton scattering in IIB superstring theory is examined in the context 
of S-duality and symmetry.  There is an algebra that generates all of 
the terms in the four-point function to any order in derivatives.  A map 
from the algebra to the scattering is given; it suggests the 
correctness of the full four-point function with the S-duality.  
The higher point functions are expected to follow a similar pattern.

\setcounter{page}{1}
\newpage
\setcounter{footnote}{0}

\noindent{\it Introduction} 

Derivative corrections to the IIB superstring low-energy effective action 
at the four-point function have been investigated in many works 
\cite{Green1}-\cite{Russo:1997mk} .  Complete formal  
perturbative results are known up to the genus two level, due to 
the complicated nature of the integrals involved 
\cite{Bao:2003im}-\cite{D'Hoker:2001nj}.

There are also several conjectures for the full four-point function, 
including instantons \cite{Chalmers1}-\cite{Green1}.  The original 
conjecture of \cite{Russo:1998vt} based on the Eisenstein functions 
failed to agree with genus one perturbation theory.  The conjecture in 
\cite{Chalmers1} was ambiguous up to relative coefficients of the 
pairings of non-holomorphic Eisenstein functions; this seems to be 
straightened out by imposing a differential condition on the modular 
construction, in \cite{Green1}. 

In this work, the systematics of the full four-point function in the 
currently accepted conjecture is examined.  The organization of the 
perturbative corrections is given in an organized manner, and illustrates 
some symmetry that is unknown in the superstring.  The four-point function 
can be found by expanding the function,  

\bqr 
\prod_{n=1}^\infty {1\over (1-2 x^{2n+1})} \ ,    
\fqr  
which is also known to be very close to certain vertex algebras.  Similar 
functions are conjectured to generate the higher-point functions.  The 
fact that the conjectured form of the amplitude can be found by expanding 
such a simple partition function appears to support its validity.  The 
form of the modular ansatz is reviewed only briefly in this work.

\vskip .2in 
\noindent{\it Brief Review} 

The low-energy effective action at four-point consists of an infinite 
number of terms, 

\bqr 
S=\int d^{10}x \sqrt{g}\Bigl[ R + {1\over\Box} R^4 + \sum_{k=0}^\infty 
 g_k(\tau,\bar\tau) \Box^k R^4 \Bigr] \ ,   
\label{fourpointaction}  
\fqr 
with $\tau=i/g_s+\theta/2\pi$, the string coupling constant, and 
$\alpha'$ has been suppressed. 
The action in \rf{fourpointaction} does not include the terms which 
contain the unitarity cuts.  The coefficients $g_k$ can be computed 
in string perturbation theory, and is difficult to obtain beyond 
genus two.  

Imposing modular invariance, in the Einstein frame of the metric, 
requires that the functions $g_k$ be invariant under fractional linear 
transformations.  Finding the functions $g_k$ at the four-point level 
has been investigated in many papers, with consistency checks.  The 
refined conjecture is investigated in this work, with the coefficients 
$g_k$ proportional to $Z_{\left\{ s \right\}}$ functions.  The latter 
functions obey a specific differential equation, and are roughly 
proportional to products of Eisenstein functions.  This ansatz is 
investigated here, and with an emphasis on extracting further symmetry.   

\vskip .2in 
\noindent{\it Tree Amplitudes}  

The four-point graviton tree amplitude has the form, 

\bqr 
A_4={64\over stu} \prod_{n=1}^\infty e^{2\zeta(2n+1)/2n/4^{2n+1}\times 
  (s^{2n+1}+t^{2n+1}+u^{2n+1})} 
\label{fourpoint}  
\fqr  
and has the expansion, 

\bqr  
{64 \over stu} \prod_i [{2\zeta(2n_i+1)\over (2n_i+1)4^{2n_i+1}} 
 (s^{2n_i+1}+t^{2n_i+1}+u^{2n_i+1})]^{m_i}~ \prod {1\over (m_i)!} \ .   
\label{fourpointexpansion}
\fqr 
The zeta functions take values in the odd integers excluding unity. 
In order to find the kinematic structure of the invidual terms, an 
identity is required that expands the Mandelstam invariants in an 
appropriate basis.  For example, 

\bqr  
s^3+t^3+u^3=3stu \ .  
\fqr 
The further higher moments are expanded as 

\bqr 
s^{2n+1}+t^{2n+1}+u^{2n+1}= c^s_n \times (stu) s^{n_s} \ , 
\fqr 
pertaining to the tensors with the maximal number of $s=(k_1+k_2)^2$ 
invariants.  In general the expansion takes the form, 

\bqr 
(s^{2n+1}+t^{2n+1}+u^{2n+1})^m \rightarrow 
  c^{n_s n_t n_u}_{n,m} \times (stu) s^{n_s} t^{n_t} u^{n_u} \ . 
\label{genkinexp}  
\fqr
The coefficients $c^{n_s n_t n_u}_n$ are found by kinematic identities 
and are in the basis in which the symmetry of $s\leftrightarrow t$, etc, 
is manifest.  

The term with $n_s\neq 0$ and $n_t=n_u=0$ is examined.  In this kinematical 
configuration, the contributions are, 

\bqr 
A_4^{n_s,0,0} = 64 \prod_i {1\over(m_i)!} 
  \left[{2\zeta(2n_i+1)\over (2n_i+1) 4^{2n_i+1}}\right]  
 c_{n,m}^{n_s} s^{n_s} \ ,
\fqr  
which at a particular order in $s$ is 

\bqr 
=64 c_{n,m}^{n_s} s^{n_s}
 \prod_i {1\over (m_i)!} \left[{2\over (2n_i+1)4^{2n_i+1}}\right]^{m_i} 
 \prod_i \zeta(2n_i+1)^{m_i} \ . 
\label{maxsexp}
\fqr 
The individual kinematic contributions follow from 1) taking a term 
$stu\times s^{n_s}$ for an integer $n_s$, then 2) partitioning the number 
$n_s+3=N$ into odd numbers $(2n_j+1)m_j$ with $m_i=1,\ldots$ so that 
$\sum m_i=m$.  The numbers $m_i$ count the 
duplicates of the numbers $2n_i+1$, in which case there are $m_i$ of the 
identical numbers for any $i$.  

The coefficient of the 
\bqr  
(stu)^m s^{n_s}  \qquad m=\sum m_i  \qquad n_s+3=\sum (2n_i+1)m_i 
\fqr 
is found by collecting the coefficient 

\bqr 
64 {1\over (m_i)!} {2\over (2n_i+1) 4^{2n_i+1}} \prod_i \zeta(2n_i+1)  \ , 
\fqr 
and multiplying by the group theory factor $c_{n,m}^{n_s}$.  The only 
constraint in this kinematic configuration is that $n_s+3=\sum (2n_i+1)m_i$.  
All possible combinations of $n_i$ an integer and $m_i$ an integer are 
allowed.  

The ansatz indicates that the zeta functions are 
to be replaced in the manner, 

\bqr  
\prod_j \zeta(2p_j+1) \rightarrow Z_{\left\{p_j+1/2\right\}} \ , 
\fqr 
with the $Z$ functions described in the next section.  

\vskip .2in 
\noindent{\it Perturbative Modular Contributions}  

The functions $Z_{\left\{ p_j+1/2\right\} }$ are described by the modular 
invariant differential equation on the torus, 

\bqr 
{1\over 4} \Delta Z_{\left\{ q_j \right\} } = 
 A Z_{ \left\{ q_j \right\} } + B \prod_j Z_{q_j} \ , 
\fqr 
with the simplest case being the Eisenstein functions, 

\bqr 
Z_s=E_s  \qquad s(s-1) Z_s = \Delta Z_s \ . 
\fqr 
The Laplacian takes the form, when restricted to the perturbative sector, 
that is, without the $\tau_1$ dependence, 

\bqr 
\Delta=4\tau_2^2\del_\tau\del_{\bar\tau} \ .  
\fqr 
The condition on $A$ and $B$ could in principle be determined generically 
by the tree and one-loop contributions of the usual perturbative string 
amplitude; however, their numbers are left unknown for the moment.

One possible set of values is 

\bqr 
B=-(\sum p_j+{1\over 2})(-1+\sum p_j+{1\over 2}) 
\fqr 
\bqr 
A=4*f*\prod (p_j+{1\over 2}) 
 \qquad f^{-1} = {1\over 4} \sum^i p_j+{1\over 2} \quad 
   {\rm or}\quad  4=2^i \ . 
\fqr 
Their form is found by computing tree and one-loop amplitudes generically, 
and there is no real reason to believe this ansatz for $A$ and $B$ 
is correct.  

The differential solution to the perturbative sector of the modular 
construction is generated as follows.  The Eisenstein functions have 
the expansion,  

\bqr 
 Z_s|_{\rm pert} = 2\zeta(2s)\tau^s + 
  {2\sqrt\pi\Gamma(s-1/2)\zeta(2s-1)\over\Gamma(s)} \tau_2^{1-s} \ .  
\fqr 
Using their form in the differential equation, 

\bqr 
 (\tau_2^2\partial^2_{\tau_2}-A) Z_{\left\{p_j+1/2\right\}} 
\fqr 
\bqr 
 = (\tau_2^2\partial^2_{\tau_2}-A) (a_0 \tau_2^{3/2+k} + ... 
  +  a_{\rm gmax} \tau_2^{3/2+k-2{\rm gmax} } ) 
\fqr 
\bqr 
 = a_0 [(3/2+k)(3/2+k-1)-A] \tau_2^{3/2+k} + ... 
\fqr 
\bqr 
 + a_{\rm gmax} [(3/2+k-2{\rm gmax})(3/2+k-2{\rm gmax}-1)-A] 
\tau_2^{3/2+k-2{\rm gmax}} 
\fqr 
\bqr 
 = b_0 \tau_2^{3/2+k} + ... + b_{\rm gmax} \tau_2^{3/2+k-2{\rm gmax}}   
\fqr 
and ignores two additional terms on the left hand side, one of which has the 
$\tau_2$ dependence to be not physical.  The number $k$ is defined by 
$\sum 2p_j+1=3/2+k$ and corresponds to the $\Box^k R^4$ term in the low-energy 
effective action.  The other one is ignored for unphysical reasons.  The 
solution to the terms is, 
 
\bqr 
a_i = {b_i\over [(3/2+k-2i)(3/2+k-2i-1)-A]}  \ . 
\fqr 
Then, the coefficients $b_i$ are found by expanding the product 
$B\times\prod_j Z_{ p_j+1/2 }$, 

\bqr 
B \prod_j Z_{p_j+1/2 } = B \prod_j 
  (a_{p_j+1/2} \tau_2^{2p_j+1} + a_{-p_j+1/2} \tau_2^{-2p_j}) \ . 
\fqr 
Examples are listed below.  For $\sum 2p_j+1=3/2+k$, 

\bqr 
b_0 = B \prod_j a_{p_j+1/2}  
\fqr 
\bqr 
a_0 = B \prod_j {a_{p_j+1/2}\over [(3/2+k)(3/2+k-1)-A]}  \ .   
\fqr 
For $\sum 2p_j+1=3/2+k-2$, 

\bqr 
b_1 = B \sum_{j\neq i} a_{p_i-1/2} \prod_j a_{p_j+1/2} 
\fqr 
\bqr 
a_1 = B \sum_{j\neq i} a_{p_i-1/2} \prod_j {a_{p_j+1/2} \over  
   [(3/2+k-2)(3/2+k-2-1)-A]}  \ .
\fqr 
The sum extends until $3/2+k-2{\rm gmax}$, in which case we have 
$\sum 2p_j+1=3/2+k-2{\rm gmax}$, 

\bqr 
b_{\rm gmax} = B \prod_j a_{-p_j+1/2}  
\fqr 
\bqr 
a_{\rm gmax} = B \prod_j {a_{-p_j+1/2}\over [(3/2+k-2{\rm gmax}) 
  (3/2+k-2{\rm gmax}-1)-A]} \ .  
\fqr 
The genus number ${\rm gmax}= {1\over 2}(2k+1)$ or ${1\over 2}(2k+2)$ 
for $k=n/2$ 
with $n$ either odd or even, and for the $\Box^k R^4$ term; all of the 
products of the $Z_{p_j+1/2}$ have a perturbative truncation due to the 
individual expansion of the $Z_{p_j+1/2}$ functions. The examples describe 
the perturbative contributions from the modular functions $Z_{\{ 2p_j+1\}}$.

\vskip .2in
\noindent{\it Quantum Extension and Symmetry}  

In this section a set of quantum rules is defined that generates the graviton
amplitudes.  The partitions of numbers are useful in parameterizing these 
contributions; also these partitions are connected to a fundamental symmetry of
the quantum theory.  

There have been several proposals for the quantum completion of the 
S-matrix, and higher derivative terms up to genus two have been computed.  
The modular invariant completion due to S-duality enforces certain 
structures on the coupling dependence.  A basis for the coupling structure 
is formed from the Eisenstein functions, the contribution of which have 
recently been elucidated more completely in \cite{Green1}.  

The polynomial system generating the perturbative contributions 
can be determined from a graphical illustration and also through 
a 'vertex' algebra.  The latter can be found from expanding the 
function, 

\bqr 
\prod_{n=1} {1\over (1-2 x^{2n+1})} \ , 
\label{modularexp}
\fqr   
which is similar to the partition function of a boson on the 
torus, 

\bqr 
\prod_{n=0} {1\over (1-x^{2n+1})} \ .  
\fqr 
The latter is associated to a vertex algebra.  The former 
will be shown to correspond to the perturbative four-point function, 
without the non-analytic terms required by unitarity.

There are a set of trees as depicted in figure 1.  Each tree is 
found by taking a number $N$, an odd number, and partitioning it 
into odd numbers 
$3,5,7,\ldots$.  At each pair of nodes the numbers in the partition 
are attached.  The partition is labeled by the set $N(\{\zeta\})$.  

\begin{figure}
\begin{center}
\epsfxsize=12cm
\epsfysize=12cm
\epsfbox{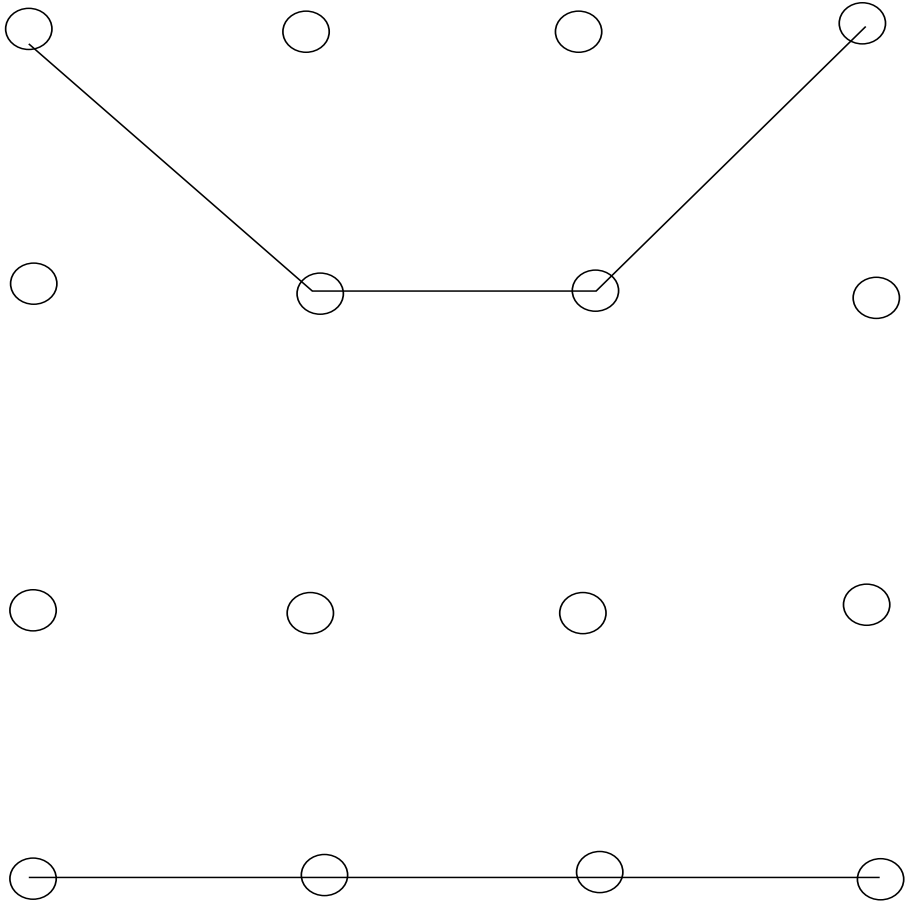}
\end{center}
\caption{Weighted trees.}
\end{figure}

The nodes of the trees are chosen, one from each pair, and a set 
of lines could be drawn between these nodes.  For each tree there are  

\bqr 
{a!\over b!(b-a)!}  
\label{treenumber}
\fqr 
ways of partitioning the tree into the same number of 'up' and 'down' 
nodes.  Each tree set is labeled by $a=N(\{\zeta\})$, and the terms 
in \rf{treenumber} are spanned by $b=0,...,N(\{\zeta\})$ for 
$a=N(\{\zeta\})$.  There are $2^{N(\{\zeta\})}$ terms or polynomials 
in each tree, as found by summing \rf{treenumber}.     

The derivative terms are labeled by $\Box^{2k}R^4$ with $k$ either 
half integral or integral starting at $k=0$.  At a specific order 
$\Box^{2k}$, the tree system is found by partitioning the number 
$N=2k+3$.  The number of partitions a number $N$ can have into odd 
numbers, excluding unity, is denoted $P_{\rm odd}(N)$.  The perturbative 
contributions to this derivative order are found as follows, 

1) attach weights to the nodes of the tree 

2) take the product of the weights of the nodes in the tree 

3) add the sums of the products 

\noindent The sum of the products reintroduces the contribution of the 
modular ansatz to the four-point scattering.  
  
For reference, the genus truncation property holds at $g_{\rm max}= 
{1\over 2}(2k+1), {1\over 2}(2k+2)$ for $k=n/2$ with $n$ either odd or even.  
There are $P_{\rm odd}(2k+3)$ partitions of $2k+3$ into odd numbers excluding 
unity, with a maximum number.  The genus truncation follows from the 
fact that there a maximum number of nodes in the tree.

Some examples of the partitions of numbers at a given derivative order 
$\Box^{2k} R^4$ are described in the following table.  For example, at 
$k=4$, the number $11$ can be partitioned into $11$ and $3+3+5$, with 
counts of $2^1$ and $2^3$.  

\bqr   
\pmatrix{ 
      k=0    &  2         &     k=4     & 2+2^3=10 \cr 
      k=1    &  2         &     k=5     & 2+2^3+2^3=18   \cr 
      k=2    &  2         &     k=6     & 2+2^5+2^3+2^3+2^3=68 \cr 
      k=3    &  2+2^3=10  &     k=7     & 2+2^3+2^5+2^3+2^3=58 \cr 
}
\fqr     
One would like to represent the terms group theoretically with the  
quantum numbers being zeta entries.  This can be done using the 
expansion in \rf{modularexp}.  

The perturbative contribution to the order $\Box^{2k}$ can be 
read off of the tree by associating the weights to the nodes.  
For the 'up' node there is a factor 

\bqr 
{2\zeta(2s)\over [(3/2+k)(3/2+k-1)-A]}, 
\fqr 
and for the 'down' node there is 

\bqr 
{{2\sqrt\pi\Gamma(s-1/2)\zeta(2s-1)\over\Gamma(s) 
[(3/2+k-2{\rm gmax})(3/2+k-2{\rm gmax}-1)-A]}} \ .  
\fqr 
Each perturbative contribution is found by multiplying the 
node contributions.  There are the various weighted trees that 
contribute to the perturbative contribution at $2k+3$, when 
partitioned into the various odd numbers.  Each contribution 
has the weighted factor of $B$ in the product.  (The weighted trees 
resemble a fermionic system with gmax fermions with a quantum level 
degeneracy non-identical fermions.)  The partitioning of the 
number $2k+3$ into the weighted trees is a convenient way of 
describing all of the contributions to the particular derivative 
term.

\vskip .2in 
\noindent{\it Partitions}

Basically the perturbative contributions to the four-point function 
come about from partitioning an integer $2k+3$ into odd numbers 
$3,5,7,\ldots$, and then subsequently choosing one of two choices 
for each number in the partition.  

The partition of a number into odd numbers can be achieved by 
expanding the function 

\bqr 
O(N)=\prod_{n=0} {1\over (1-x^{2n+1})}  
\fqr 
into the polynomials 

\bqr 
x^{n_1} x^{n_2} \ldots x^{n_m} \ , 
\fqr 
for all sets of numbers $n_i$.  The expansion and summation of 
all terms generates the sum $\sum P(N) x^N$ with $P(N)$ the total number 
of partitions.  

The number of partitions of the number $N$ into all odd numbers, and 
excluding unity, is achieved by expanding the function 

\bqr 
O1(N)=\prod_{n=1} {1\over (1-x^{2n+1})} \ .  
\fqr 
This expansion generates the sum $\sum P1(N) x^N$ with $P1(N)$ numbering 
the total number of partitions, with 

\bqr 
P1(N)={1\over N!} \partial_x^N ~\prod_{n=1}^\infty {1\over (1-x^{2n+1})} \ . 
\fqr 
The sum $O1(N)$ follows from the same expansion as $O(N)$ but with 
additional factor of $(1-x)$ removed from the denominator in the 
infinite product.   

In the CFT language, the function $O(N)$ arises from a scalar on 
the torus.  The $O1(N)$ function has one mode deleted, and represents 
a non-modular modification.  

To complete the count of the number of tree systems the factor of two 
must be added to each of the nodes.  The number of trees is, 

\bqr 
T(N)=\sum_{\rm tree ~systems} 2^{N(\{ \zeta \} )} \ ,    
\fqr 
with the corresponding sum without the $2^{N({\zeta})}$ being the 
count $P1(N)$.  This number $T(N)$ can be found from, 

\bqr 
P1(N)={1\over N!} \partial_x^N ~ \prod_{n=1}^\infty {1\over (1-2x^{2n+1})} \ , 
\fqr
and the $2$ at each node follows from the number of $x$s in the 
expansion, in all possible partitions.  The relevant function is 

\bqr 
\prod_{n=1}^\infty {1\over (1-2 x^{2n+1})} \ , 
\fqr 
and in the cft language requires another modification of the modes 
by a factor of $2$.  Termwise the rescaling of $x$ to $2x$ corresponds 
with $x=e^{-2\pi \tau_2}$, with $\tau_2\rightarrow \tau_2- 
\ln 2^{-1/2\pi}$.  

\vskip .2in 
\noindent{\it Higher-Point Functions} 

The higher point functions require expanding the function $\sqrt{g} \Box^{2k} 
R^4$ to higher order in the plane wave expansion, which stems essentially 
from the $\sqrt{g}$.  There are also terms with higher numbers of curvatures, 
requiring $R^p$ at $p$-point.  The tree amplitudes of the higher point 
functions are not currently avalable in the literature, in either an 
expanded or closed form.  

One conjecture for the higher derivative terms at $p$-point would be 
to take the form, 
\bqr 
64c_{n,m}^{n_s} s^{n_s} 
 \prod_i {1\over (m_i)!} \left[{2\over (2n_i+1)4^{2n_i+1}}\right]^{m_i} 
 \prod_i \zeta(2n_i+1)^{m_i} \ , 
\fqr 
with the $\sum 2n_i+1=5,7,\ldots$.  The sum indicates that there are 
this number of vertices in a tree-level $p$-point graph, such as $p=5$ 
with $\sum 2n_i+1=5$ on up; likewise, at $p=7$ the sum starts at $9$.  
The tensor combination here is for the terms with the maximum number 
of $s$ invariants, and the tensor function $c_{n,m}^{n_s}$ pertains 
to the $p$-point.  Also, the helicity structure is found from the 
particular combination in the collection of Weyl tensors.  

The origin of the $\zeta(3)$ in the $R^4$ stems from the fact that 
there are two vertices, with a massive propagator between them.  In 
the low-energy limit of the string amplitude, this results in an 
infinite sum of the massive modes, with a $1/n$ coming from each 
vertex and also the propagator.  At higher point, the sum of the 
massive modes requires more vertices, and results in the tree-level 
$\zeta$ function starting out at $2p-5$, which is the number of vertices 
and propagators in a skeleton graph; this graph is a ladder tree diagram 
with $m-4$ external gravitons on the internal rung.  

Internal gravitons within the tree diagram enforce the entire tree 
diagram to be built of gravitons.  This is in the absence of a 
massive-graviton-graviton vertex (or massive-graviton ... graviton 
vertex).  Of course, at four-point we know there is a 
massive-graviton-graviton vertex, with a factor of $1/n$ from the 
vertex.  The ladder diagrams result in the local terms in the 
amplitudes, that is without any $1/\Box$ terms in the low energy 
effective action.  And each replacement of the massive mode by 
a graviton eliminates two factors of $1/n$ in the infinite sum; 
this lowers the $\zeta$ function value by two.    

Assuming the $\zeta$ function takes on values in the tree diagrams 
at $p$-point beginning at $2p-5$, it is natural to think that the 
partition function, 

\bqr 
\sum_{i=0}^{p+3}~ \prod_{n=2p-5-2i} {1\over (1-2x^{2n+1})} \ , 
\fqr 
would be able to generate these scattering amplitudes.  The helicity 
tensor from the $R^{2p}$ and the tensor function $c_{n,m}^{n_s,\ldots}$ 
is required.  Without more information this is just a partial 
conjecture for the form of the amplitude, however, the partition function 
seems appropriate.  The terms with $i\neq 1$ correspond to diagrams 
without the maximal number of internal massive modes; these diagrams 
produce potential $1/\Box$ terms in the low-energy effective action.  
(For example, at four-point, there is a contribution ${1\over\Box} R^4$ 
contribution from the GR graphs.)

\vskip .3in 
\noindent{\it Conclusion}

The systematics of the four-point amplitude according to the S-duality 
of the IIB superstring is explored and delimited.  The systematics 
entail the partitioning of numbers and the construction of weighted 
trees.  The entire four-point function, without the non-analytic 
terms, can be found through a simple construction.  

The procedure of finding the four-point function is found by expanding 
a particular partition function.  This partition function should be 
related to the S-duality of the IIB superstring.  The simplicity of 
the generating function, 

\bqr 
\prod_{i=1} {1\over (1-2x^{2n+1})} \ , 
\fqr 
does seem to suggest the correctness of the four-point function.  The 
higher point functions have similar partition functions.   

The form of the amplitude can be used to deduce contributions in 
the massless sector, that is maximal supergravity.  These calculations 
are typically very complicated, especially at the multi-loop level.  
The techniques for doing these cancellations are improving, and 
possible cancellations leading to better ultra-violet behavior 
beyond those up to five loops have been investigated in 
\cite{Bern:1998ug}-\cite{Chalmers2}.

The origin of the simple partition function that generates the scattering 
seems not clear.  There could be some additional symmetry beyond the naive 
supersymmetry and S-duality, or an extension as a result of the two.  Some 
indication of this is examined in \cite{Damour:2005zb}.  Vertex algebras 
are generated by these partition functions \cite{VertexAlgebra}.

\vfill\break

\end{document}